\documentclass[11pt]{article}

\usepackage[utf8]{inputenc}
\usepackage[english]{babel}
\usepackage[a4paper, left=1in, right=1in, top=1.25in, bottom=1.25in]{geometry}
\usepackage{graphicx}
\usepackage{float}
\usepackage{amsmath,amssymb,amsthm,mathtools,mathrsfs}
\usepackage{physics}
\usepackage{enumitem}
\usepackage{hyperref}
\usepackage{authblk}
\usepackage[toc,page,title,titletoc,header]{appendix}

\newcommand{\C}{\mathbb{C}}

\newcommand{\E}{\mathbb{E}}
\newcommand{\1}{\mathbf{1}}
\newcommand{\MK}[1]{\mathcal{M}_{#1}(\C)}
\newcommand{\re}{\mathrm{Re}\,}

\newcommand{\gfrak}{\mathfrak{g}}
\newcommand{\ofrak}{\mathfrak{o}}
\newcommand{\Qmat}{\mathbf{Q}}
\newcommand{\Htr}{\mathcal{H}}
\newcommand{\Rtr}{\mathcal{R}}

\newcommand{\TrN}{\operatorname{Tr}_N}
\newcommand{\TrK}{\operatorname{Tr}_K}
\newcommand{\ptrN}{\mathcal{T}_N}

\theoremstyle{definition}
\newtheorem{definition}{Definition}[section]
\newtheorem{result}{Result}[section]

\newtheorem*{definition*}{\textit{Definition}}
\newtheorem*{result*}{\textit{Result}}

\newenvironment{remarks}{%
  \par\vspace{1ex}
  \noindent\textbf{Remarks.}\begin{itemize}\setlength\itemsep{2pt}}%
  {\end{itemize}\par\vspace{1ex}}

\title{$\mathcal{R}$-transforms for non-Hermitian block random matrices:\\ a spherical integral approach}

\author[1,2]{Pierre Bousseyroux\thanks{Email: pierre.bousseyroux@polytechnique.edu}}

\affil[1]{Econophysics Lab, Institut Louis Bachelier, 28 Pl.\ de la Bourse, Palais Brongniart, 75002 Paris, France}
\affil[2]{LadHyX, UMR CNRS 7646, \'Ecole polytechnique, Institut Polytechnique de Paris, 91128 Palaiseau, France}

\date{}

\begin{document}

\maketitle

\begin{abstract}
We extend the spherical-integral approach to $\mathcal{R}$-transforms of non-Hermitian random matrices, recently developed in \cite{BousseyrouxPotters2026R}, to random matrices with a fixed $K\times K$ block structure. We introduce a rank-$K$ spherical integral defining a scalar $\mathcal{H}$-transform on $2K\times 2K$ overlap matrices and an associated matrix-valued $\mathcal{R}$-transform encoding the joint block structure. Using the replica method, we derive a conjectural subordination relation for sums of independent block matrices. This framework provides a method for determining the spectral boundaries of large non-Hermitian random matrices with a fixed block structure.
\end{abstract}

\section{Introduction}
\label{sec:intro}

Random matrices with a fixed block structure form a broad class of inhomogeneous random matrix models. They can be written as
\[
\vb{M}
=
\bigl(\vb{M}_{ab}\bigr)_{1\leq a,b\leq K},
\]
where the number of blocks $K$ remains fixed while the dimensions of the individual blocks grow. Throughout the paper we restrict, for simplicity, to blocks of equal size $N\times N$, with $K$ fixed as $N\to\infty$. A classical subclass is formed by finite-block variance-profile ensembles: the index set is partitioned into $K$ macroscopic groups of comparable size, and the entries are independent, with distributions---or at least variances---depending only on the groups to which their row and column indices belong. Such matrices arise naturally whenever a large disordered system contains a finite number of interacting populations, agents, or internal variables. Examples include multi-species neural networks, where stability and the transition to chaos are governed by the spectrum of a block-structured connectivity matrix \cite{AljadeffSternSharpee2015,AljadeffRenfrewStern2015}, MIMO communication channels \cite{RashidiFarOrabyBrycSpeicher2008}, and stability matrices arising in ecological and economic systems \cite{PatilAguirreLopezBouchaud2024}.

In the Hermitian setting, block random matrices have been studied through canonical equations and operator-valued free probability. Girko's canonical equations provided early self-consistent descriptions of ensembles with structured variances \cite{Girko2001Canonical}, while explicit models with large random blocks were investigated by Oraby \cite{Oraby2007SpectralLaws,Oraby2007Girko}. From the operator-valued point of view, the finite block structure survives in the large-$N$ limit as a finite-dimensional algebra over which conditional expectations are taken \cite{Shlyakhtenko1996,RashidiFarOrabyBrycSpeicher2006,RashidiFarOrabyBrycSpeicher2008}. The corresponding operator-valued Cauchy transforms satisfy matrix-valued self-consistent equations \cite{HeltonRashidiFarSpeicher2007}; a detailed account is given in \cite{MingoSpeicher2017}.

For non-Hermitian finite-block variance profiles, Aljadeff, Renfrew and Stern derived implicit equations for the limiting density and an explicit formula for the spectral radius \cite{AljadeffRenfrewStern2015}. More general deterministic variance profiles were subsequently studied by Cook, Hachem, Najim and Renfrew \cite{CookHachemNajimRenfrew2018,CookHachemNajimRenfrew2022}, while matrix-Dyson-equation methods were developed for non-Hermitian Kronecker random matrices, including correlated block structures \cite{AltErdosKrugerNemish2019}. More recently, Patil, Aguirre-L\'opez and Bouchaud considered jointly Gaussian block matrices with correlations between distinct blocks and derived matrix-valued equations for their spectral boundaries \cite{PatilAguirreLopezBouchaud2024}. Their framework was illustrated on stability matrices arising in ecological and economic models.

In the unstructured case $K=1$, a spherical-integral framework for non-Hermitian $\mathcal{R}$-transforms was recently proposed in \cite{BousseyrouxPotters2026R}. There, the two scalar functions $\Rtr_1$ and $\Rtr_2$ governing the addition of free non-Hermitian matrices \cite{FeinbergZee1997Addition,JanikNowakPappZahed1997,BurdaJanikNowak2011} were shown to arise as derivatives of a single scalar function $\Htr$. This function is defined through a spherical integral generalizing the Harish-Chandra--Itzykson--Zuber integral \cite{HarishChandra1957,ItzyksonZuber1980,GuionnetMaida2005,BenaychGeorges2011}. 

Here we extend this construction to random matrices with a fixed $K\times K$ block structure. The rank-$K$ transform is defined from the joint block structure under simultaneous unitary conjugation and does not require the different blocks to be independent. Blockwise rotational invariance is imposed only in the independent-block specialization of Section~\ref{sec:blockwise}, where the joint transform reduces to scalar transforms of the individual blocks. Our main contributions are the following.
\begin{itemize}
\setlength{\itemsep}{2pt}
    \item We define a \emph{rank-$K$ spherical integral}, namely a scalar functional $\Htr^{K}_{\vb{M}}$ of a $2K\times 2K$ overlap matrix, which reduces, for $K=1$, to the scalar construction of \cite{BousseyrouxPotters2026R}, with the explicit ensemble average just described when the matrix is random (Section~\ref{sec:rankK}). Its matrix derivatives define a matrix-valued $\mathcal{R}$-transform and satisfy an additive relation under independent Haar block conjugation.
    \item Using the replica method, we derive a conjectural rank-$K$ \emph{subordination relation} for block matrices satisfying
    \[
        \vb{M}_{ij}
        =
        \vb{A}_{ij}+\vb{V}\vb{B}_{ij}\vb{V}^{*},
        \qquad 1\le i,j\le K,
    \]
    where $\vb{A}$ and $\vb{B}$ are independent and the same Haar unitary $\vb{V}$ acts on every block. The relation is stated in Section~\ref{sec:result} and derived formally in Appendix~\ref{app:replica}.
    \item When the blocks of $\vb{B}$ are \emph{independent and rotationally invariant}, we show that the joint transform simplifies: $\Htr^{K}_{\vb{B}}$ decomposes into the sum of the scalar $\Htr$-transforms of the individual blocks (Result~\ref{res:blockwise}). For independent Gaussian blocks, the corresponding fixed-point equations reduce to the known self-consistent equations of finite-block variance-profile models \cite{AljadeffRenfrewStern2015,CookHachemNajimRenfrew2018}.
    \item The rank-$K$ subordination relation yields matrix-valued fixed-point equations that provide a systematic method for determining the spectral boundaries of large non-Hermitian random matrices with a fixed block structure (Section~\ref{sec:boundaries}).
    \item We illustrate the formalism through the block linearization of the product
    \[
        (\1+\vb{X}_1)(\1+\vb{X}_2)
    \]
    of two shifted elliptic matrices, and compare the predicted spectral boundaries with numerical simulations (Section~\ref{sec:example}).
\end{itemize}

\section{Scalar case: notation and main relation}
\label{sec:scalar}

We briefly recall the notation and the scalar relation of \cite{BousseyrouxPotters2026R}. Let $\vb{M}$ be an $N\times N$ matrix and let $z,\omega\in\C$. Define the $2N\times 2N$ Hermitized matrix
\begin{equation}
    \mathcal{C}_{\vb{M}}(\omega,z)
    :=
    \begin{pmatrix}
        \omega\,\1_N & z\1_N-\vb{M}\\[2pt]
        \bar z\1_N-\vb{M}^{*} & \omega\,\1_N
    \end{pmatrix},
    \qquad
    \vb{G}_{\vb{M}}(\omega,z)
    :=
    \mathcal{C}_{\vb{M}}(\omega,z)^{-1}
    =
    \begin{pmatrix}
        \vb{G}_{11} & \vb{G}_{12}\\
        \vb{G}_{21} & \vb{G}_{22}
    \end{pmatrix},
\end{equation}
where $\vb{M}^{*}$ denotes the conjugate transpose. The normalized block traces
\begin{equation}
    \gfrak_{ab}^{N}(\omega,z)
    :=
    \frac{1}{N}\TrN \vb{G}_{ab}(\omega,z),
    \qquad a,b\in\{1,2\},
\end{equation}
are collected into
\begin{equation}
    \mathcal{G}_{\vb{M}}^{N}(\omega,z)
    :=
    \begin{pmatrix}
        \gfrak_{11}^{N} & \gfrak_{12}^{N}\\
        \gfrak_{21}^{N} & \gfrak_{22}^{N}
    \end{pmatrix},
    \qquad
    \mathcal{G}_{\vb{M}}
    :=
    \lim_{N\to\infty}\mathcal{G}_{\vb{M}}^{N}.
\end{equation}
We write $\gfrak_{ab}$ for the entries of $\mathcal{G}_{\vb{M}}$. The two diagonal normalized traces coincide, $\gfrak_{11}=\gfrak_{22}=:\gfrak_1$, and we write $\gfrak_2:=\gfrak_{21}$, explicitly
\begin{align}
    \gfrak_{1}(\omega,z)
    &=
    \lim_{N\to\infty}\frac{1}{N}\TrN\!\left[
        \omega\bigl(\omega^{2}\1-(z\1-\vb{M})(z\1-\vb{M})^{*}\bigr)^{-1}
    \right],
    \\
    \gfrak_{2}(\omega,z)
    &=
    -\lim_{N\to\infty}\frac{1}{N}\TrN\!\left[
        (z\1-\vb{M})^{*}\bigl(\omega^{2}\1-(z\1-\vb{M})(z\1-\vb{M})^{*}\bigr)^{-1}
    \right].
\end{align}
In the limit $\omega\to0$ with $z$ outside the spectrum, $\gfrak_2(\omega,z)$ converges to the normalized Cauchy transform
\[
g(z)=\lim_{N\to\infty}\frac{1}{N}\TrN\bigl[(z\1-\vb{M})^{-1}\bigr].
\]

Following \cite{BousseyrouxPotters2026R}, the central object is a spherical integral of the following form. For random ensembles we explicitly include the ensemble average; this is the convention used below for independent random blocks.

\begin{definition*}
Let $\vb{M}$ be a deterministic or random matrix of size $N\times N$, and let $\psi_1,\psi_2\in\C^{N}$. We define the functional $\Htr_{\vb{M}}^{N}$ by
\begin{equation}\label{eq:H_scalar}
\frac{1}{2N}
\log
\E_{\vb{M}}\E_{\vb{U}}
\left[
\exp\left(
2N\,\re
\langle
\psi_1,\vb{U}\vb{M}\vb{U}^{*}\psi_2
\rangle
\right)
\right]
=
\Htr_{\vb{M}}^{N}
\left(
\|\psi_1\|\,\|\psi_2\|,
\langle\psi_1,\psi_2\rangle,
\overline{\langle\psi_1,\psi_2\rangle}
\right),
\end{equation}
where $\vb{U}$ is Haar-distributed on $\mathrm{U}(N)$, the unitary group of $N\times N$ matrices, and $\langle x,y\rangle=\sum_i \bar x_i y_i$. The expectation $\E_{\vb{M}}$ is omitted when $\vb{M}$ is deterministic. We then set
\[
\Htr_{\vb{M}}
=
\lim_{N\to\infty}
\Htr_{\vb{M}}^{N}.
\]
The associated scalar transforms are
\begin{equation}\label{eq:R_scalar}
\Rtr_1(\alpha,\beta,\bar\beta) := \partial_\alpha \Htr(\alpha,\beta,\bar\beta),
\qquad
\Rtr_2(\alpha,\beta,\bar\beta) := 2\,\partial_\beta \Htr(\alpha,\beta,\bar\beta).
\end{equation}
If $\beta=x+iy$, we use the Wirtinger derivatives
\[
\partial_{\beta}
=
\frac12\left(\partial_x-i\partial_y\right),
\qquad
\partial_{\bar\beta}
=
\frac12\left(\partial_x+i\partial_y\right),
\]
treating $\beta$ and $\bar\beta$ as independent variables for differentiation. When the arguments arise from a pair of vectors, $\alpha\ge0$ and the third argument is the complex conjugate of the second; in that case we abbreviate $\Rtr_i(\alpha,\beta,\bar\beta)$ as $\Rtr_i(\alpha,\beta)$. We denote by $\widetilde{\Rtr}_1,\widetilde{\Rtr}_2$ the multivalued analytic functions collecting all determinations (branches) of $\Rtr_1,\Rtr_2$.
\end{definition*}

If $\vb{M}=\vb{A}+\vb{W}\vb{B}\vb{W}^{*}$ with $\vb{A},\vb{B}$ independent and $\vb{W}$ an independent Haar unitary, then $\Htr_{\vb{M}}=\Htr_{\vb{A}}+\Htr_{\vb{B}}$, hence the $\Rtr_i$ are additive. The main result of \cite{BousseyrouxPotters2026R} is the subordination relation
\begin{equation}\label{eq:conj_scalar}
    \mathcal{G}_{\vb{A}+\vb{W}\vb{B}\vb{W}^{*}}(\omega,z)
    =
    \mathcal{G}_{\vb{A}}
    \bigl(
        \omega-\Rtr_{1,\vb{B}}(\gfrak_1,\gfrak_2),\;
        z-\Rtr_{2,\vb{B}}(\gfrak_1,\gfrak_2)
    \bigr),
\end{equation}
with $\gfrak_i=\gfrak_{i,\vb{A}+\vb{W}\vb{B}\vb{W}^{*}}(\omega,z)$, valid for large $\omega$ and analytically continued elsewhere with an appropriate choice of branches. We then set $\omega=-i\varepsilon$ with $\varepsilon>0$ and let $\varepsilon\downarrow0$. One has $\gfrak_1\to i\,\ofrak(z)$ with $\ofrak(z)\geq0$; the function $\ofrak$ is nonzero exactly on the support of the limiting spectral density $\rho$, and is related to the Chalker--Mehlig eigenvector self-overlap \cite{ChalkerMehlig1998,JanikEtAl1999}. This limit is the key to the spectral-boundary equations of \cite{bousseyroux2026boundaries, bousseyroux2026multiplicative}: edges are reached when $\ofrak(z)\to 0^{+}$.

\section{Block matrices and rank-\texorpdfstring{$K$}{K} spherical integrals}
\label{sec:rankK}

\subsection{Setting and notation}

We now consider $K\times K$ block matrices
\begin{equation}
    \vb{M} = \bigl(\vb{M}_{ij}\bigr)_{1\le i,j\le K}\in \mathcal{M}_K\bigl(\mathcal{M}_N(\C)\bigr),
\end{equation}
where $K$ is fixed and $N\to\infty$. We use $\TrN$ and $\TrK$ for traces over the $N$- and $K$-dimensional indices, respectively. To keep the block structure explicit, for any scalar $d\times d$ matrix $S$ we denote by
\begin{equation}\label{eq:block_lift}
    [S]_N
    :=
    \bigl(S_{ij}\1_N\bigr)_{1\le i,j\le d}
\end{equation}
its block lift. For a $d\times d$ block matrix $\mathcal{X}=(\mathcal{X}_{ij})$ with $N\times N$ blocks, we also define the normalized partial trace
\begin{equation}\label{eq:partial_trace}
    \bigl(\ptrN[\mathcal{X}]\bigr)_{ij}
    :=
    \frac{1}{N}\TrN \mathcal{X}_{ij}.
\end{equation}
We now allow the scalar variable $z$ to be replaced by a matrix $Z\in\MK{K}$. For general
\begin{equation}
    \Omega
    =
    \begin{pmatrix}
        W_1 & Y_{12}\\
        Y_{21} & W_2
    \end{pmatrix}
    \in \MK{2K},
\end{equation}
we define the $2NK\times 2NK$ block matrix
\begin{equation}\label{eq:hermitization_block}
    \mathcal{C}_{\vb{M}}(\Omega)
    :=
    \begin{pmatrix}
        [W_1]_N & [Y_{12}]_N-\vb{M}\\[2pt]
        [Y_{21}]_N-\vb{M}^{*} & [W_2]_N
    \end{pmatrix}
\end{equation}
and its inverse
\begin{equation}
    \vb{G}_{\vb{M}}(\Omega)
    :=
    \mathcal{C}_{\vb{M}}(\Omega)^{-1}
    =
    \begin{pmatrix}
        \vb{G}_{11} & \vb{G}_{12}\\
        \vb{G}_{21} & \vb{G}_{22}
    \end{pmatrix},
\end{equation}
each $\vb{G}_{ab}$ being itself a $K\times K$ block matrix with $N\times N$ blocks. Taking normalized partial traces defines
\begin{equation}
    \gfrak_{ab,\vb{M}}(\Omega)
    :=
    \lim_{N\to\infty}\ptrN\bigl[\vb{G}_{ab}(\Omega)\bigr],
    \qquad a,b\in\{1,2\},
\end{equation}
collected into
\begin{equation}
    \mathcal{G}_{\vb{M}}(\Omega)
    :=
    \begin{pmatrix}
        \gfrak_{11,\vb{M}}(\Omega) & \gfrak_{12,\vb{M}}(\Omega)\\
        \gfrak_{21,\vb{M}}(\Omega) & \gfrak_{22,\vb{M}}(\Omega)
    \end{pmatrix}
    \in\MK{2K}.
\end{equation}
For
\begin{equation}
    \Omega(\omega,Z):=\begin{pmatrix}\omega\1_K & Z\\ Z^{*} & \omega\1_K\end{pmatrix}
\end{equation}
with real positive $\omega$ sufficiently large, $\mathcal{C}_{\vb{M}}$ is Hermitian positive definite. The formulas are subsequently continued to complex $\omega$. For $K=1$ we recover the objects of Section~\ref{sec:scalar}, with $\gfrak_{11}=\gfrak_1$ and $\gfrak_{21}=\gfrak_2$.

\subsection{The rank-\texorpdfstring{$K$}{K} \texorpdfstring{$\Htr$}{H}-transform}

Let $\phi_{a,1},\dots,\phi_{a,K}\in\C^{N}$ for $a\in\{1,2\}$, and form the $N\times K$ matrices $\Phi_a=(\phi_{a,1},\dots,\phi_{a,K})$. Define the overlap matrix
\begin{equation}\label{eq:overlap_def}
    \Gamma
    =
    \begin{pmatrix}
        \Gamma_{11} & \Gamma_{12}\\
        \Gamma_{21} & \Gamma_{22}
    \end{pmatrix}
    \in\MK{2K},
    \qquad
    (\Gamma_{ab})_{ij}
    :=
    \langle\phi_{b,j},\phi_{a,i}\rangle.
\end{equation}
For $\vb{V}\in\mathrm{U}(N)$, let $\mathscr{C}_{\vb{V}}$ denote simultaneous conjugation of all blocks,
\begin{equation}\label{eq:block_conjugation}
    \bigl[\mathscr{C}_{\vb{V}}(\vb{M})\bigr]_{ij}
    :=
    \vb{V}\,\vb{M}_{ij}\,\vb{V}^{*}.
\end{equation}
As in the scalar case, the same Haar unitary $\vb{V}$ acts on every block.

\begin{definition}[Rank-$K$ $\Htr$-transform]\label{def:HK}
Let $\vb{M}$ be a deterministic or random block matrix. For $\Phi_1,\Phi_2\in\C^{N\times K}$, define
\begin{equation}\label{eq:HK_def}
    \Htr^{N,K}_{\vb{M}}(\Gamma)
    :=
    \frac{1}{2N}
    \log
    \E_{\vb{M}}\E_{\vb{V}}
    \left[
    \exp
    \Bigl(
        2N \re \sum_{i,j=1}^{K}
        \bigl\langle \phi_{1,i},\,
        \bigl[\mathscr{C}_{\vb{V}}(\vb{M})\bigr]_{ij}\,
        \phi_{2,j} \bigr\rangle
    \Bigr)
    \right],
\end{equation}
where $\vb{V}$ is Haar-distributed on $\mathrm{U}(N)$ and $\E_{\vb{M}}$ is omitted when $\vb{M}$ is deterministic. Whenever the limit exists, we set $\Htr^{K}_{\vb{M}}:=\lim_{N\to\infty}\Htr^{N,K}_{\vb{M}}$. The associated matrix $\Rtr$-transform is defined, for $X\in\MK{2K}$, by
\begin{equation}\label{eq:RK_def}
    \bigl[\Rtr^{K}_{\vb{M}}(X)\bigr]_{ab,ij}
    :=
    2\left.
    \frac{\partial\Htr^{K}_{\vb{M}}}{\partial(\Gamma_{ba})_{ji}}
    \right|_{\Gamma=X},
    \qquad a,b\in\{1,2\}.
\end{equation}
The derivatives are understood entrywise in the Wirtinger sense. We write
\[
\Rtr^{K}_{\vb{M}}
=
\begin{pmatrix}
\Rtr^{K}_{11,\vb{M}} &
\Rtr^{K}_{12,\vb{M}}\\
\Rtr^{K}_{21,\vb{M}} &
\Rtr^{K}_{22,\vb{M}}
\end{pmatrix}.
\]
We denote by $\widetilde{\Rtr}^{K}_{\vb{M}}$ the multivalued analytic function collecting all determinations of $\Rtr^{K}_{\vb{M}}$, and use the corresponding analytic branches when needed.
\end{definition}

The quantity in \eqref{eq:HK_def} depends on the vectors only through the overlap matrix $\Gamma$. Indeed, two families $\{\phi_{a,i}\}$ with the same overlap matrix differ by a global unitary rotation, which can be absorbed into the Haar matrix $\vb{V}$. This is the direct rank-$K$ analogue of the scalar reduction in Section~\ref{sec:scalar}. If $\vb{M}$ is itself \emph{block-rotationally invariant}, in the sense that
\begin{equation}\label{eq:invariance}
    \mathscr{C}_{\vb{V}}(\vb{M})
    \overset{\mathrm{law}}{=}
    \vb{M}
    \qquad\text{for all } \vb{V}\in\mathrm{U}(N),
\end{equation}
then the Haar average in \eqref{eq:HK_def} may be omitted, leaving only the ensemble average over $\vb{M}$, as in the scalar rotationally invariant setting of \cite{BousseyrouxPotters2026R}.

\begin{remarks}
    \item \textbf{Additivity.} If $\vb{M}=\vb{A}+\mathscr{C}_{\vb{V}}(\vb{B})$ with $\vb{A},\vb{B}$ independent and $\vb{V}$ Haar, then the exponent in \eqref{eq:HK_def} is linear in $\vb{M}$, so that
    \begin{equation}
        \Htr^{K}_{\vb{M}} = \Htr^{K}_{\vb{A}} + \Htr^{K}_{\vb{B}},
        \qquad\text{hence}\qquad
        \Rtr^{K}_{\vb{M}} = \Rtr^{K}_{\vb{A}} + \Rtr^{K}_{\vb{B}},
    \end{equation}
    with the same relation at the level of multivalued extensions.
    \item \textbf{Reduction to $K=1$.} For $K=1$, $\beta=\langle\phi_1,\phi_2\rangle=\Gamma_{21}$. For an argument $X$, let
    \[
    \alpha=\sqrt{X_{11}X_{22}},
    \qquad
    \beta=X_{21}.
    \]
    Equation~\eqref{eq:RK_def} gives
    \[
    \Rtr^{1}_{11}
    =
    \sqrt{\frac{X_{22}}{X_{11}}}\,
    \Rtr_1(\alpha,\beta),
    \qquad
    \Rtr^{1}_{22}
    =
    \sqrt{\frac{X_{11}}{X_{22}}}\,
    \Rtr_1(\alpha,\beta),
    \]
    and
    \[
    \Rtr^{1}_{12}
    =
    \Rtr_2(\alpha,\beta).
    \]
    When evaluated on the scalar resolvent, $X_{11}=X_{22}=\gfrak_1$ and $X_{21}=\gfrak_2$. Hence
    \[
    \Rtr^{1}_{11}
    =
    \Rtr^{1}_{22}
    =
    \Rtr_1(\gfrak_1,\gfrak_2),
    \qquad
    \Rtr^{1}_{12}
    =
    \Rtr_2(\gfrak_1,\gfrak_2),
    \]
    in agreement with the scalar convention of Section~\ref{sec:scalar}.
    \item \textbf{Scaling relation.} The exponent in \eqref{eq:HK_def} is invariant under $\Phi_1\mapsto\lambda\Phi_1$, $\Phi_2\mapsto\bar\lambda^{-1}\Phi_2$, $\lambda\in\C^{*}$. Hence
    \begin{equation}
        \Htr^{K}\bigl(|\lambda|^{2}\Gamma_{11},\,|\lambda|^{-2}\Gamma_{22},\,\Gamma_{12},\,\Gamma_{21}\bigr)
        =
        \Htr^{K}(\Gamma),
    \end{equation}
    which generalizes the fact that, for $K=1$, $\Htr$ depends on $\|\psi_1\|,\|\psi_2\|$ only through the product $\|\psi_1\|\|\psi_2\|$.
\end{remarks}

\section{The rank-\texorpdfstring{$K$}{K} subordination relation}
\label{sec:result}

We can now state the main result of this paper. It is obtained through the replica method; the derivation, which follows the scalar one of \cite{BousseyrouxPotters2026R} with matrix-valued order parameters, is presented in Appendix~\ref{app:replica}.

\begin{result}\label{conj:main}
Let $\vb{A}$ and $\vb{B}$ be two large independent block matrices, let $\vb{V}$ be Haar-distributed on $\mathrm{U}(N)$, and write $\vb{M}=\vb{A}+\mathscr{C}_{\vb{V}}(\vb{B})$, i.e. $\vb{M}_{ij}=\vb{A}_{ij}+\vb{V}\vb{B}_{ij}\vb{V}^{*}$ for every pair of block indices $(i,j)$. Assume that the large-$N$ block resolvent self-averages with respect to the Haar rotation. The replica calculation then gives, for $\omega$ large enough,
\begin{equation}\label{eq:conj_main}
\mathcal{G}_{\vb{M}}\bigl(\Omega(\omega,Z)\bigr)
=
\mathcal{G}_{\vb{A}}
\left(
\Omega(\omega,Z)
-
\Rtr^{K}_{\vb{B}}
\left(
\mathcal{G}_{\vb{M}}\bigl(\Omega(\omega,Z)\bigr)
\right)
\right).
\end{equation}
The relation is then analytically continued away from this regime, with the appropriate branch of $\widetilde{\Rtr}^{K}_{\vb{B}}$.
\end{result}

Blockwise, Eq.~\eqref{eq:conj_main} states that the two diagonal blocks $\omega\1_K$ are shifted by $\Rtr^{K}_{11,\vb{B}}$ and $\Rtr^{K}_{22,\vb{B}}$, while $Z$ and $Z^{*}$ are shifted by $\Rtr^{K}_{12,\vb{B}}$ and $\Rtr^{K}_{21,\vb{B}}$, respectively. All four blocks are evaluated at the block resolvent of $\vb{M}$. This is the direct analogue of the scalar result \eqref{eq:conj_scalar}: no a priori rotational invariance of $\vb{B}$ is assumed, the addition being implemented by the same Haar conjugation on every block.

\begin{remarks}
\item In the present non-Hermitian setting, the use of analytic
continuation of the $\mathcal R$-transforms, together with their
different determinations, was introduced in
\cite{BousseyrouxPotters2026R}.  This point of view provides a convenient
way of expressing spectral results directly in terms of suitable
branches of these transforms.
See
\cite{bousseyroux2026boundaries,bousseyroux2026outliers,bousseyroux2026multiplicative,BousseyrouxPotters2026MultiplicativeFiniteRank,Bousseyroux2026Overlaps}.
    \item \textbf{Closed equations for block-scalar $\vb{A}$.} Suppose $\vb{A}=[a]_N$ with $a\in\MK{K}$, i.e. $\vb{A}_{ij}=a_{ij}\1_N$. Then every block of $\mathcal{C}_{\vb{A}}(\Omega)$ is proportional to $\1_N$, so that $\mathcal{G}_{\vb{A}}(\Omega)=\bigl(\Omega-\mathcal{A}\bigr)^{-1}$ exactly, where
    \begin{equation}
        \mathcal{A} := \begin{pmatrix}0 & a\\ a^{*} & 0\end{pmatrix}.
    \end{equation}
    Result~\ref{conj:main} then closes into the $2K\times 2K$ fixed-point equation
    \begin{equation}\label{eq:fixed_point}
        \mathcal{G}
        =
        \left(
            \Omega(\omega,Z) - \mathcal{A} - \Rtr^{K}_{\vb{B}}(\mathcal{G})
        \right)^{-1}
    \end{equation}
    This is the fixed-point equation used in the examples below. It is a $2K\times2K$ matrix equation, equivalently a system for four $K\times K$ blocks.
    \item \textbf{Consistency with $K=1$.} For $K=1$ and $\vb{A}$ arbitrary, \eqref{eq:conj_main} reduces to the scalar result \eqref{eq:conj_scalar}.
\end{remarks}

\section{Blockwise independent ensembles}
\label{sec:blockwise}

The rank-$K$ transform simplifies when the blocks are independent.

\begin{result}\label{res:blockwise}
Let $\vb{B}=(\vb{B}_{ij})$ be a block matrix whose blocks are independent, each $\vb{B}_{ij}$ being rotationally invariant ($\vb{V}\vb{B}_{ij}\vb{V}^{*}\overset{\mathrm{law}}{=}\vb{B}_{ij}$ for all $\vb{V}\in\mathrm{U}(N)$). Then
\begin{equation}\label{eq:H_decomposition}
    \Htr^{K}_{\vb{B}}(\Gamma)
    =
    \sum_{i,j=1}^{K}
    \Htr_{\vb{B}_{ij}}
    \bigl(
        \alpha_{ij},\,
        \beta_{ij},\,
        \bar\beta_{ij}
    \bigr),
    \qquad
    \alpha_{ij} := \sqrt{(\Gamma_{11})_{ii}\,(\Gamma_{22})_{jj}},
    \quad
    \beta_{ij} := (\Gamma_{21})_{ji},
    \quad
    \bar\beta_{ij} := (\Gamma_{12})_{ij},
\end{equation}
where $\Htr_{\vb{B}_{ij}}$ is the scalar $\Htr$-transform \eqref{eq:H_scalar} of the block $\vb{B}_{ij}$.
\end{result}

\begin{proof}
The simultaneously conjugated blocks are $[\mathscr{C}_{\vb{V}}(\vb{B})]_{ij}=\vb{V}\vb{B}_{ij}\vb{V}^{*}$. Recall that for random $\vb{B}$ Definition~\ref{def:HK} includes the ensemble average over its blocks. The exponent in \eqref{eq:HK_def} is therefore a sum over the pairs $(i,j)$,
\begin{equation}
    \sum_{i,j=1}^{K}
    2N\re
    \bigl\langle
        \phi_{1,i},\,
        \vb{V}\vb{B}_{ij}\vb{V}^{*}\,\phi_{2,j}
    \bigr\rangle.
\end{equation}
Independence factorizes the expectation over the blocks into a product of $K^{2}$ factors. Since each block is rotationally invariant, each factor is independent of the common Haar matrix $\vb{V}$; the remaining Haar average is therefore trivial. One obtains
\begin{equation}
    \Htr^{N,K}_{\vb{B}}(\Gamma)
    =
    \sum_{i,j}
    \frac{1}{2N}\log\E\Bigl[\exp\bigl(2N\re\langle\phi_{1,i},\vb{B}_{ij}\phi_{2,j}\rangle\bigr)\Bigr].
\end{equation}
Each term is precisely the defining integral \eqref{eq:H_scalar} of the scalar $\Htr$-transform of $\vb{B}_{ij}$, evaluated on the pair of vectors $(\phi_{1,i},\phi_{2,j})$. By rotational invariance of $\vb{B}_{ij}$, this quantity depends only on
\begin{equation}
    \|\phi_{1,i}\|\,\|\phi_{2,j}\|
    =
    \sqrt{(\Gamma_{11})_{ii}\,(\Gamma_{22})_{jj}}
    =
    \alpha_{ij}
\end{equation}
and on $\langle\phi_{1,i},\phi_{2,j}\rangle=(\Gamma_{21})_{ji}=\beta_{ij}$ together with its conjugate $(\Gamma_{12})_{ij}=\bar\beta_{ij}$. Taking the limit $N\to\infty$ term by term yields \eqref{eq:H_decomposition}.
\end{proof}

Evaluating the preceding quantities at $\Gamma=\mathcal{G}$, we have
\[
\alpha_{ij}
=
\sqrt{(\mathcal{G}_{11})_{ii}(\mathcal{G}_{22})_{jj}},
\qquad
\beta_{ij}
=
(\mathcal{G}_{21})_{ji},
\qquad
\bar\beta_{ij}
=
(\mathcal{G}_{12})_{ij}.
\]
Differentiating \eqref{eq:H_decomposition} and using \eqref{eq:RK_def} gives
\begin{equation}\label{eq:R11_blockwise}
    \bigl(\Rtr^{K}_{11,\vb{B}}\bigr)_{ii}
    =
    \sum_{j=1}^{K}
    \sqrt{
    \frac{(\mathcal{G}_{22})_{jj}}
         {(\mathcal{G}_{11})_{ii}}
    }
    \,
    \Rtr_{1,\vb{B}_{ij}}(\alpha_{ij},\beta_{ij},\bar\beta_{ij}),
\end{equation}
\begin{equation}
    \bigl(\Rtr^{K}_{22,\vb{B}}\bigr)_{jj}
    =
    \sum_{i=1}^{K}
    \sqrt{
    \frac{(\mathcal{G}_{11})_{ii}}
         {(\mathcal{G}_{22})_{jj}}
    }
    \,
    \Rtr_{1,\vb{B}_{ij}}(\alpha_{ij},\beta_{ij},\bar\beta_{ij}),
\end{equation}
and
\begin{equation}\label{eq:R12_blockwise}
    \bigl(\Rtr^{K}_{12,\vb{B}}\bigr)_{ij}
    =
    \Rtr_{2,\vb{B}_{ij}}(\alpha_{ij},\beta_{ij},\bar\beta_{ij}).
\end{equation}
Similarly,
\begin{equation}\label{eq:R21_blockwise}
    \bigl(\Rtr^{K}_{21,\vb{B}}\bigr)_{ji}
    =
    2\,\partial_{\bar\beta}
    \Htr_{\vb{B}_{ij}}
    \bigl(\alpha_{ij},\beta_{ij},\bar\beta_{ij}\bigr).
\end{equation}

\begin{remarks}
    \item \textbf{Gaussian blocks: recovering variance profiles.} Take $\vb{B}_{ij}=\sigma_{ij}\vb{X}_{ij}$ with $\vb{X}_{ij}$ independent complex Ginibre matrices. The scalar transforms are $\Htr(\alpha,\beta)=\sigma_{ij}^{2}\alpha^{2}/2$, $\Rtr_1=\sigma_{ij}^{2}\alpha$, $\Rtr_2=0$ \cite{BousseyrouxPotters2026R}. Equations \eqref{eq:R11_blockwise}--\eqref{eq:R12_blockwise} give
    \begin{equation}
        \bigl(\Rtr^{K}_{11}\bigr)_{ii}=\sum_j \sigma_{ij}^{2}\,(\mathcal{G}_{22})_{jj},
        \qquad
        \bigl(\Rtr^{K}_{22}\bigr)_{jj}=\sum_i \sigma_{ij}^{2}\,(\mathcal{G}_{11})_{ii},
        \qquad
        \Rtr^{K}_{12}=0,
    \end{equation}
    and \eqref{eq:fixed_point} reduces to the self-consistent equations of the block variance model, first obtained in \cite{AljadeffRenfrewStern2015} and rigorously developed in \cite{CookHachemNajimRenfrew2018,AltErdosKruger2018Circular}.
    \item \textbf{Bi-invariant blocks.} If every block $\vb{B}_{ij}$ is bi-invariant\footnote{$\vb{B}_{ij}$ and $\vb{U}\vb{B}_{ij}\vb{V}$ have the same law for all unitary $\vb{U},\vb{V}$.}, then $\Rtr_{2,\vb{B}_{ij}}=0$ and $\Rtr_{1,\vb{B}_{ij}}$ is the $R$-transform of the symmetrized singular-value distribution of the block \cite{BousseyrouxPotters2026R,HaagerupLarsen2000,GuionnetKrishnapurZeitouni2011}. For bi-invariant blocks, $\Rtr_2=0$, and the fixed-point equation depends on each block only through $\Rtr_1$.
    \item \textbf{Correlated blocks.} For correlated blocks the factorization of Result~\ref{res:blockwise} no longer holds, but Definition~\ref{def:HK} still applies to the joint block law. In the jointly Gaussian case, the rank-$K$ transform remains explicit because the corresponding exponential moment is Gaussian.
\end{remarks}

\section{Spectral boundaries and the matrix \texorpdfstring{$\Qmat(z)$}{Q(z)}}
\label{sec:boundaries}

We now set $Z=z\1_K$ and study the limit $\omega\to0$, which contains the spectral information. As in the scalar case, the correct prescription inside the spectrum is to set $\omega=-i\varepsilon$ with $\varepsilon>0$ and let $\varepsilon\downarrow0$. In this limit we define the $K\times K$ matrix
\begin{equation}\label{eq:Q_def}
\Qmat(z)
:=
\lim_{\varepsilon\downarrow0}
(-i)\,
\gfrak_{11,\vb{M}}
\bigl(
\Omega(-i\varepsilon,z\1_K)
\bigr).
\end{equation}
We also define
\[
\mathbf{g}(z)
:=
\lim_{\varepsilon\downarrow0}
\gfrak_{21,\vb{M}}
\bigl(
\Omega(-i\varepsilon,z\1_K)
\bigr),
\qquad
g(z)
:=
\frac1K\TrK\mathbf{g}(z).
\]
Then $g(z)$ is the normalized Cauchy transform of the limiting eigenvalue measure. Whenever this measure admits a density,
\[
\rho(z)
=
\frac1\pi
\partial_{\bar z}g(z).
\]

The matrix $\Qmat(z)$ is Hermitian positive semi-definite and generalizes the scalar order parameter $\ofrak(z)$ of \cite{BousseyrouxPotters2026R,bousseyroux2026boundaries}. For the ensembles considered below, and assuming a regular two-dimensional limiting density, $\TrK\Qmat(z)$ is positive in the interior of the spectral support and tends to zero when a regular boundary point is approached from inside. We therefore use the vanishing of $\TrK\Qmat(z)$ as a criterion for locating the spectral boundary. In the scalar setting, the corresponding order parameter is related to eigenvector-overlap observables.

Numerically, the fixed-point equation is solved at $\omega=-i\varepsilon$ with a small $\varepsilon>0$, and the boundary is extracted from the vanishing of this order parameter.

\section{Example: product of two shifted elliptic matrices}
\label{sec:example}

Block linearization can be used to study polynomial functions of random matrices \cite{MingoSpeicher2017}. We consider the product
\begin{equation}
\label{eq:shifted_product}
\vb{P}
:=
(\1+\vb{X}_1)(\1+\vb{X}_2),
\end{equation}
where $\vb{X}_1$ and $\vb{X}_2$ are independent $N\times N$ complex Gaussian elliptic matrices with parameter $\tau\in(-1,1)$ \cite{Girko1986Elliptic,SommersEtAl1988}. Our convention is
\begin{equation}
\E\bigl|(X_k)_{ij}\bigr|^{2}=\frac{1}{N},
\qquad
\E(X_k)_{ij}(X_k)_{ji}=\frac{\tau}{N},
\qquad i\neq j,
\end{equation}
for $k=1,2$, with different unordered pairs independent. The precise diagonal convention is asymptotically irrelevant here. For $\tau=0$ one recovers the complex Ginibre ensemble. We restrict to $|\tau|<1$, for which the limiting spectrum is genuinely two-dimensional; the cases $\tau=\pm1$ are degenerate limits.

We linearize \eqref{eq:shifted_product} by introducing the $2N\times2N$ matrix
\begin{equation}
\label{eq:linearization_model}
    \vb{M}
    =
    \begin{pmatrix}
        0 & \1+\vb{X}_1\\
        \1+\vb{X}_2 & 0
    \end{pmatrix}
    =
    \underbrace{\begin{pmatrix}0&\1\\ \1&0\end{pmatrix}}_{\vb{A}\,=\,[a]_N}
    +
    \underbrace{\begin{pmatrix}0&\vb{X}_1\\ \vb{X}_2&0\end{pmatrix}}_{\vb{B}},
    \qquad
    a=\begin{pmatrix}0&1\\1&0\end{pmatrix}.
\end{equation}
Thus $K=2$ throughout this example. We have
\begin{equation}
    \vb{M}^{2}
    =
    \begin{pmatrix}
        (\1+\vb{X}_1)(\1+\vb{X}_2) & 0\\
        0 & (\1+\vb{X}_2)(\1+\vb{X}_1)
    \end{pmatrix},
\end{equation}
so the eigenvalues of $\vb{M}$ are the two square roots of the eigenvalues of $\vb{P}$. We write $\zeta$ for an eigenvalue of $\vb{M}$ and $z=\zeta^{2}$ for the corresponding eigenvalue of $\vb{P}$.

We now apply the rank-$2$ fixed-point equation to $\vb{M}$. Since $K=2$, we write
\begin{equation}
\mathcal{G}
=
\begin{pmatrix}
\gfrak_{11} & \gfrak_{12}\\
\gfrak_{21} & \gfrak_{22}
\end{pmatrix}
\in\MK{4},
\qquad
\gfrak_{ab}\in\MK{2}.
\end{equation}
For notational simplicity in this example, we also write $\Rtr_{ab}:=\Rtr^{2}_{ab,\vb{B}}$.

The two non-zero random blocks of $\vb{B}$ are independent copies of the elliptic ensemble. Their scalar transform is
\begin{equation}
\Htr(\alpha,\beta,\bar\beta)
=
\frac{\alpha^{2}}{2}
+
\frac{\tau}{4}
\bigl(\beta^{2}+\bar\beta^{2}\bigr),
\end{equation}
and hence
\begin{equation}
\Rtr_1(\alpha,\beta,\bar\beta)=\alpha,
\qquad
\Rtr_2(\alpha,\beta,\bar\beta)=\tau\beta,
\end{equation}
as in \cite{BousseyrouxPotters2026R}. Result~\ref{res:blockwise} therefore gives
\begin{equation}
    \Rtr_{11}
    =
    \begin{pmatrix}
        (\gfrak_{22})_{22} & 0\\
        0 & (\gfrak_{22})_{11}
    \end{pmatrix},
    \qquad
    \Rtr_{22}
    =
    \begin{pmatrix}
        (\gfrak_{11})_{22} & 0\\
        0 & (\gfrak_{11})_{11}
    \end{pmatrix},
\end{equation}
together with
\begin{equation}
    \Rtr_{12}
    =
    \tau
    \begin{pmatrix}
        0 & (\gfrak_{21})_{21}\\
        (\gfrak_{21})_{12} & 0
    \end{pmatrix},
    \qquad
    \Rtr_{21}
    =
    \tau
    \begin{pmatrix}
        0 & (\gfrak_{12})_{21}\\
        (\gfrak_{12})_{12} & 0
    \end{pmatrix}.
\end{equation}

The fixed-point equation \eqref{eq:fixed_point}, evaluated at $Z=\zeta\1_2$, is therefore
\begin{equation}\label{eq:example_fixed_point}
    \mathcal{G}
    =
    \begin{pmatrix}
        \omega-(\gfrak_{22})_{22} & 0 & \zeta & -1-\tau(\gfrak_{21})_{21}\\[2pt]
        0 & \omega-(\gfrak_{22})_{11} & -1-\tau(\gfrak_{21})_{12} & \zeta\\[2pt]
        \bar\zeta & -1-\tau(\gfrak_{12})_{21} & \omega-(\gfrak_{11})_{22} & 0\\[2pt]
        -1-\tau(\gfrak_{12})_{12} & \bar\zeta & 0 & \omega-(\gfrak_{11})_{11}
    \end{pmatrix}^{-1}.
\end{equation}

The spectral boundary can be extracted from this equation. At a regular boundary point approached from outside, $\omega=0$ and the diagonal blocks vanish,
\begin{equation}
\gfrak_{11}=\gfrak_{22}=0.
\end{equation}
The exchange symmetry between the two identically distributed factors selects the symmetric solution
\begin{equation}
\label{eq:g21_uv}
\gfrak_{21}
=
\begin{pmatrix}
u & v\\
v & u
\end{pmatrix},
\qquad
\gfrak_{12}
=
\gfrak_{21}^{*}.
\end{equation}
Substituting \eqref{eq:g21_uv} into \eqref{eq:example_fixed_point} at $\omega=0$ gives
\begin{equation}
\begin{pmatrix}
u & v\\
v & u
\end{pmatrix}
=
\begin{pmatrix}
\zeta & -1-\tau v\\
-1-\tau v & \zeta
\end{pmatrix}^{-1},
\end{equation}
or equivalently
\begin{equation}
\label{eq:uv_holomorphic}
u
=
\frac{\zeta}
{\zeta^{2}-(1+\tau v)^{2}},
\qquad
v
=
\frac{1+\tau v}
{\zeta^{2}-(1+\tau v)^{2}}.
\end{equation}

With $z=\zeta^{2}$, the second equation in \eqref{eq:uv_holomorphic} becomes
\begin{equation}
\label{eq:z_of_v}
z=z_{\tau}(v)
:=
1+\tau+\frac{1}{v}
+2\tau v+\tau^{2}v^{2}.
\end{equation}
Equivalently, $v=v(z)$ solves
\begin{equation}
\label{eq:v_cubic}
\tau^{2}v^{3}
+2\tau v^{2}
+(1+\tau-z)v
+1
=0,
\end{equation}
with $v(z)\sim z^{-1}$ as $|z|\to\infty$.

Let $q_1=(\Qmat(\zeta))_{11}$ and $q_2=(\Qmat(\zeta))_{22}$. Linearizing the fixed-point equation around $\Qmat=0$ gives
\begin{equation}
\label{eq:Q_linearization_example}
\begin{pmatrix}
q_1\\
q_2
\end{pmatrix}
=
\begin{pmatrix}
|v|^{2} & |u|^{2}\\
|u|^{2} & |v|^{2}
\end{pmatrix}
\begin{pmatrix}
q_1\\
q_2
\end{pmatrix}
+o(q_1+q_2).
\end{equation}
A nontrivial solution exists when
\begin{equation}
\label{eq:uv_edge}
|u|^{2}+|v|^{2}=1.
\end{equation}
From \eqref{eq:uv_holomorphic},
\begin{equation}
\frac{u}{v}
=
\frac{\zeta}{1+\tau v},
\end{equation}
and since $|\zeta|^{2}=|z|$, this becomes
\begin{equation}
\label{eq:general_product_boundary}
|v|^{2}
\bigl(
|1+\tau v|^{2}
+
|z_{\tau}(v)|
\bigr)
=
|1+\tau v|^{2},
\qquad
z_{\tau}(v)
=
1+\tau+\frac{1}{v}+2\tau v+\tau^{2}v^{2}.
\end{equation}
Equation~\eqref{eq:general_product_boundary} gives the spectral boundary of \eqref{eq:shifted_product} for any $\tau\in(-1,1)$, with the solution of \eqref{eq:v_cubic} satisfying $v(z)\sim z^{-1}$ at infinity.

The Ginibre case $\tau=0$ provides a useful check. Then
\begin{equation}
z_{0}(v)=1+\frac{1}{v},
\end{equation}
and the boundary equation reduces to
\begin{equation}
|v|^{2}(1+|z|)=1.
\end{equation}
Since $v=(z-1)^{-1}$, this is equivalent to
\begin{equation}
|z-1|^{2}=1+|z|.
\end{equation}
Writing $z=re^{i\phi}$ gives, away from $z=0$,
\begin{equation}
\label{eq:pascal_limacon}
r=1+2\cos\phi,
\qquad
|\phi|\leq\frac{2\pi}{3}.
\end{equation}
This is the lima\c{c}on obtained previously for the product of two shifted Ginibre matrices \cite{BurdaJanikNowak2011}. The case $\tau=0$ of \eqref{eq:general_product_boundary} therefore recovers that result.

\begin{figure}[H]
    \centering
    \includegraphics[width=\textwidth]{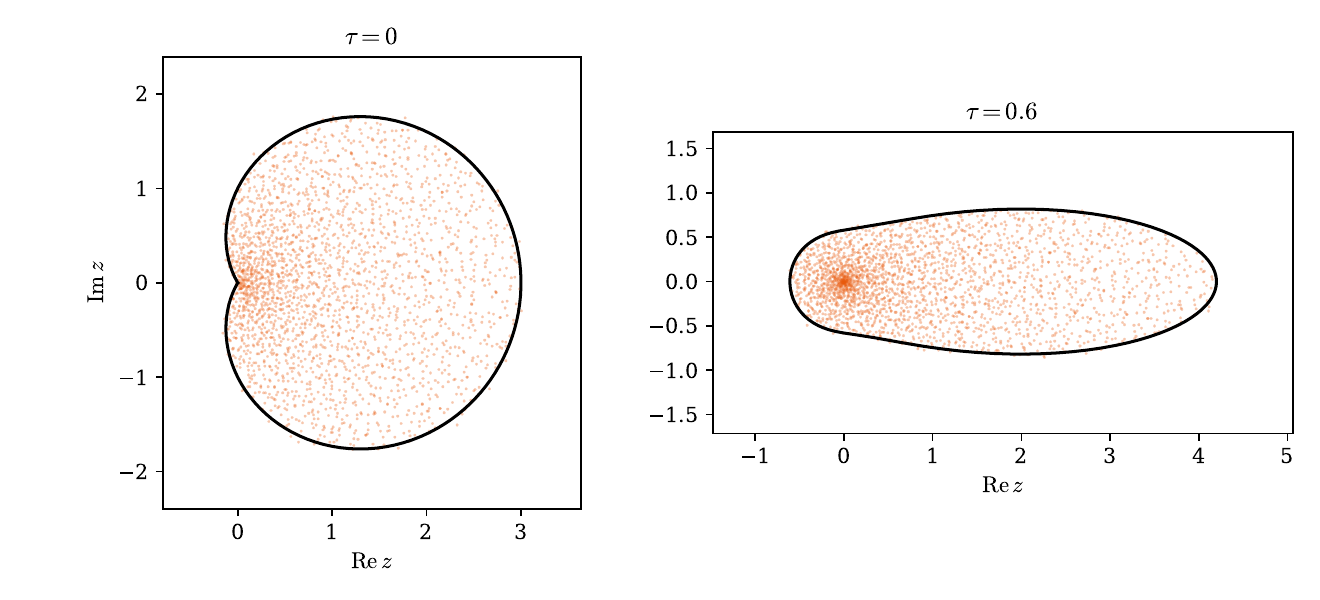}
    \caption{
    Eigenvalues of the product $\vb{P}=(\1+\vb{X}_1)(\1+\vb{X}_2)$ for $\tau=0$ (left) and $\tau=0.6$ (right). Orange points are eigenvalues from numerical realizations with $N=800$, and black curves are the limiting spectral boundaries predicted by \eqref{eq:general_product_boundary}. For $\tau=0$, the general elliptic formula reduces to the known Pascal lima\c{c}on $r=1+2\cos\phi$, $|\phi|\leq 2\pi/3$ \cite{BurdaJanikNowak2011}.
    }
    \label{fig:shifted_product}
\end{figure}

Figure~\ref{fig:shifted_product} compares \eqref{eq:general_product_boundary} directly with eigenvalues of the product $\vb{P}$ for $\tau=0$ and $\tau=0.6$.

\paragraph{Acknowledgements.}

This research was conducted within the Econophysics \& Complex Systems Research Chair, under the aegis of the Fondation du Risque, the Fondation de l'\'Ecole polytechnique, the \'Ecole polytechnique, and Capital Fund Management. Generative AI tools were used for language editing.

\bibliographystyle{plain}
\bibliography{References}

\begin{appendices}

\section{Replica derivation of Result~\ref{conj:main}}
\label{app:replica}

The derivation parallels the scalar one of \cite{BousseyrouxPotters2026R} (itself modeled on the Hermitian computation of \cite{BunAllezBouchaudPotters2016}), with scalar order parameters promoted to $K\times K$ matrices. As is customary with replicas \cite{MezardParisiVirasoro1987}, the computation is formal. Normalization factors independent of the order parameters are omitted throughout.

\subsection{Replica representation of the block resolvent}

Let $\mathcal{C}$ be a $2NK\times2NK$ Hermitian matrix with positive eigenvalues, written in the block form \eqref{eq:hermitization_block}. For $a,b\in\{1,2\}$, the exact complex Gaussian identity is the ratio
\begin{equation}\label{eq:gaussian_quotient}
    \bigl[\mathcal{C}^{-1}\bigr]_{ab}
    =
    \frac{
    \displaystyle
    \int
    \varphi_a\varphi_b^{*}
    \exp\bigl(-\langle\varphi,\mathcal{C}\,\varphi\rangle\bigr)
    \,\dd\varphi
    }{
    \displaystyle
    \int
    \exp\bigl(-\langle\varphi,\mathcal{C}\,\varphi\rangle\bigr)
    \,\dd\varphi
    },
\end{equation}
where $\varphi=(\varphi_1,\varphi_2)$ with $\varphi_a\in\C^{NK}$ and $\dd\varphi=\dd\varphi_1\,\dd\varphi_2/\pi^{2NK}$. The denominator equals $(\det\mathcal{C})^{-1}$. Introducing $n$ independent copies $\varphi^{(1)},\dots,\varphi^{(n)}$ and multiplying the numerator and the denominator by
\[
\Bigl(\int\exp\bigl(-\langle\varphi,\mathcal{C}\,\varphi\rangle\bigr)\,\dd\varphi\Bigr)^{n-1}
=
(\det\mathcal{C})^{1-n}
\]
yields
\begin{equation}
    \bigl[\mathcal{C}^{-1}\bigr]_{ab}
    =
    \frac{
    \displaystyle
    \int
    \varphi_a^{(1)}\bigl(\varphi_b^{(1)}\bigr)^{*}
    \exp\Bigl(-\sum_{r=1}^{n}\langle\varphi^{(r)},\mathcal{C}\,\varphi^{(r)}\rangle\Bigr)
    \prod_{r=1}^{n}\dd\varphi^{(r)}
    }{
    \displaystyle
    \int
    \exp\Bigl(-\sum_{r=1}^{n}\langle\varphi^{(r)},\mathcal{C}\,\varphi^{(r)}\rangle\Bigr)
    \prod_{r=1}^{n}\dd\varphi^{(r)}
    }.
\end{equation}
The denominator is $(\det\mathcal{C})^{-n}$. The formal limit $n\to0$ therefore gives
\begin{equation}\label{eq:replica_rep}
    \bigl[\mathcal{C}^{-1}\bigr]_{ab}
    =
    \lim_{n\to0}
    \int
    \varphi_a^{(1)}\bigl(\varphi_b^{(1)}\bigr)^{*}
    \exp
    \Bigl(
        -\sum_{r=1}^{n}
        \bigl\langle \varphi^{(r)},\, \mathcal{C}\,\varphi^{(r)} \bigr\rangle
    \Bigr)
    \prod_{r=1}^{n}\frac{\dd\varphi_1^{(r)}\,\dd\varphi_2^{(r)}}{\pi^{2NK}}.
\end{equation}
After averaging over the common Haar rotation, the replicas are in principle coupled through overlap matrices carrying both replica and block indices. Following the scalar calculation of \cite{BousseyrouxPotters2026R}, we impose a replica-diagonal (replica-symmetric) saddle and display one representative replica. This is the main non-rigorous step of the derivation: possible off-diagonal replica order parameters are discarded before the formal $n\to0$ continuation.

We now specialize to $\mathcal{C}=\mathcal{C}_{\vb{M}}(\Omega(\omega,Z))$. We condition on $\vb{A}$ throughout this calculation (or simply take $\vb{A}$ deterministic); if $\vb{A}$ is random, the final deterministic relation additionally uses the self-averaging assumption stated in Result~\ref{conj:main}. We write
\begin{equation}
    \vb{M}=\vb{A}+\mathscr{C}_{\vb{V}}(\vb{B}),
\end{equation}
and $\vb{V}$ Haar-distributed on $\mathrm{U}(N)$. We decompose each replica vector over the blocks, writing $\varphi_a=(\phi_{a,1},\dots,\phi_{a,K})$ with $\phi_{a,i}\in\C^{N}$, or equivalently $\Phi_a=(\phi_{a,1},\dots,\phi_{a,K})\in\C^{N\times K}$. The quadratic form then splits into
\begin{equation}
\begin{aligned}
    \bigl\langle\varphi,\mathcal{C}_{\vb{M}}\varphi\bigr\rangle
    &=
    \omega\TrK\bigl(\Phi_1^{*}\Phi_1\bigr)
    +\omega\TrK\bigl(\Phi_2^{*}\Phi_2\bigr)
    \\
    &\qquad
    +2\re
    \sum_{i,j}
    \Bigl[
        Z_{ij}\,\langle\phi_{1,i},\phi_{2,j}\rangle
        -
        \langle\phi_{1,i},\vb{M}_{ij}\,\phi_{2,j}\rangle
    \Bigr].
\end{aligned}
\end{equation}
The averaged resolvent is therefore proportional to
\begin{equation}
\begin{aligned}
    \E_{\vb{B}}\E_{\vb{V}}
    \bigl[
        \vb{G}_{ab,\vb{M}}(\Omega)
    \bigr]
    &\propto
    \int
    \Phi_a\Phi_b^{*}
    \exp
    \Bigl\{
        -\omega\TrK(\Phi_1^{*}\Phi_1)
        -\omega\TrK(\Phi_2^{*}\Phi_2)
    \Bigr\}
    \\
    &\qquad\times
    \exp
    \Bigl\{
        -2\re\sum_{i,j}Z_{ij}\,\langle\phi_{1,i},\phi_{2,j}\rangle
        +2\re\sum_{i,j}\langle\phi_{1,i},\vb{A}_{ij}\phi_{2,j}\rangle
    \Bigr\}
    \\
    &\qquad\times
    \E_{\vb{B}}\E_{\vb{V}}
    \Bigl[
        \exp
        \Bigl\{
            2\re\sum_{i,j}
            \langle\phi_{1,i},[\mathscr{C}_{\vb{V}}(\vb{B})]_{ij}\phi_{2,j}\rangle
        \Bigr\}
    \Bigr]
    \,\dd\Phi_1\,\dd\Phi_2.
\end{aligned}
\end{equation}

\subsection{Averaging over the noise and matrix order parameters}

By Definition~\ref{def:HK}, the joint average over the Haar rotation and the noise ensemble is precisely the rank-$K$ $\Htr$-transform. We introduce the normalized overlap matrices
\begin{equation}
    C_{ij}
    :=
    \frac1N\langle\phi_{1,j},\phi_{1,i}\rangle,
    \qquad
    D_{ij}
    :=
    \frac1N\langle\phi_{2,j},\phi_{2,i}\rangle,
    \qquad
    E_{ij}
    :=
    \frac1N\langle\phi_{1,j},\phi_{2,i}\rangle.
\end{equation}
Applying Definition~\ref{def:HK} to the rescaled vectors $N^{-1/2}\Phi_1$ and $N^{-1/2}\Phi_2$ gives
\begin{equation}
    \E_{\vb{B}}\E_{\vb{V}}
    \exp
    \Bigl(
        2\re\sum_{i,j}\langle\phi_{1,i},[\mathscr{C}_{\vb{V}}(\vb{B})]_{ij}\phi_{2,j}\rangle
    \Bigr)
    =
    \exp
    \Bigl(
        2N\,
        \Htr^{N,K}_{\vb{B}}
        \Bigl(
        \begin{pmatrix}
        C & E^{*}\\
        E & D
        \end{pmatrix}
        \Bigr)
    \Bigr).
\end{equation}
We enforce these order parameters by inserting matrix delta functions with Lagrange multipliers $P,Q$ (Hermitian $K\times K$) and $R\in\MK{K}$. These constraints are represented by the exponential factors below, with the multiplier contours chosen along the appropriate imaginary directions:
\begin{equation}
\begin{aligned}
    1
    &\propto
    \int
    \exp
    \Bigl\{
        \sum_{i,j}P_{ij}\,\langle\phi_{1,i},\phi_{1,j}\rangle
        -N\TrK(PC)
    \Bigr\}
    \,\dd P\,\dd C,
    \\
    1
    &\propto
    \int
    \exp
    \Bigl\{
        \sum_{i,j}Q_{ij}\,\langle\phi_{2,i},\phi_{2,j}\rangle
        -N\TrK(QD)
    \Bigr\}
    \,\dd Q\,\dd D,
    \\
    1
    &\propto
    \int
    \exp
    \Bigl\{
        2\re\sum_{i,j}R_{ij}\,\langle\phi_{1,i},\phi_{2,j}\rangle
        -2N\re\TrK(RE)
    \Bigr\}
    \,\dd R\,\dd E.
\end{aligned}
\end{equation}
After this insertion, the integrand is Gaussian in $(\Phi_1,\Phi_2)$, with the same covariance structure as the replica representation of $\mathcal{G}_{\vb{A}}$, but with shifted parameters:
\begin{equation}
    \omega\1_K\;\longmapsto\;\omega\1_K-P,
    \qquad
    \omega\1_K\;\longmapsto\;\omega\1_K-Q,
    \qquad
    Z\;\longmapsto\;Z-R .
\end{equation}
Performing the Gaussian integral therefore yields, at the saddle point,
\begin{equation}\label{eq:saddle_subordination}
    \E_{\vb{B}}\E_{\vb{V}}
    \bigl[
        \mathcal{G}_{\vb{M}}(\Omega(\omega,Z))
    \bigr]
    \;\approx\;
    \mathcal{G}_{\vb{A}}
    \begin{pmatrix}
        \omega\1_K-P & Z-R\\
        (Z-R)^{*} & \omega\1_K-Q
    \end{pmatrix},
\end{equation}
where $\approx$ has the usual formal meaning of the replica saddle-point approximation, and where $(P,Q,R,C,D,E)$ extremize the finite-dimensional action
\begin{equation}\label{eq:action}
\begin{aligned}
    \mathcal{S}
    &=
    2\,\Htr^{K}_{\vb{B}}
    \Bigl(
    \begin{pmatrix}
        C & E^{*}\\
        E & D
    \end{pmatrix}
    \Bigr)
    -\TrK(PC)-\TrK(QD)
    \\
    &\qquad
    -2\re\TrK(RE)
    +
    \mathcal{L}_{\vb{A}}
    \bigl(
        \widetilde{\Omega}
    \bigr),
\end{aligned}
\end{equation}
with
\begin{equation}\label{eq:L_A}
    \widetilde{\Omega}
    :=
    \begin{pmatrix}
        \omega\1_K-P & Z-R\\
        (Z-R)^{*} & \omega\1_K-Q
    \end{pmatrix},
    \qquad
    \mathcal{L}_{\vb{A}}(\widetilde{\Omega})
    :=
    -\lim_{N\to\infty}\frac{1}{N}\log\det\mathcal{C}_{\vb{A}}(\widetilde{\Omega}).
\end{equation}

\subsection{Saddle-point equations}

The Gaussian insertion for the resolvent block $(a,b)$ contains the dyad $\phi_{a,i}\phi_{b,j}^{*}$. Its normalized trace is
\[
\frac{1}{N}\TrN\bigl(\phi_{a,i}\phi_{b,j}^{*}\bigr)
=
\frac{1}{N}\langle\phi_{b,j},\phi_{a,i}\rangle.
\]
At the saddle point, the normalized overlap matrix is therefore
\[
\begin{pmatrix}
C & E^{*}\\
E & D
\end{pmatrix}
=
\mathcal{G}.
\]

Differentiating \eqref{eq:action} with respect to the order parameters $C,D,E$ and using \eqref{eq:RK_def} gives
\begin{equation}\label{eq:saddle_multipliers}
    P
    =
    \Rtr^{K}_{11,\vb{B}},
    \qquad
    Q
    =
    \Rtr^{K}_{22,\vb{B}},
    \qquad
    R
    =
    \Rtr^{K}_{12,\vb{B}},
\end{equation}
and, in the initial Hermitian domain,
\[
\Rtr^{K}_{21,\vb{B}}
=
2\,\partial_{E^{*}}\,\Htr^{K}_{\vb{B}}.
\]
Differentiating \eqref{eq:action} with respect to the Lagrange multipliers $P,Q,R$ and using
\begin{equation}
    \partial_t\log\det \mathcal{C}(t)
    =
    \operatorname{Tr}_{2NK}\!\bigl[\mathcal{C}(t)^{-1}\,\mathcal{C}'(t)\bigr]
\end{equation}
together with the definition \eqref{eq:L_A} of $\mathcal{L}_{\vb{A}}$ recovers the identification of the overlap variables with $\mathcal{G}$. Substituting \eqref{eq:saddle_multipliers} into \eqref{eq:saddle_subordination} therefore yields
\begin{equation}
    \E_{\vb{B}}\E_{\vb{V}}
    \bigl[
        \mathcal{G}_{\vb{M}}\bigl(\Omega(\omega,Z)\bigr)
    \bigr]
    \;\approx\;
    \mathcal{G}_{\vb{A}}
    \Bigl(
        \Omega(\omega,Z)-\Rtr^{K}_{\vb{B}}\bigl(\mathcal{G}_{\vb{M}}(\Omega(\omega,Z))\bigr)
    \Bigr).
\end{equation}

\end{appendices}

\end{document}